\documentclass[]{spie}  
\usepackage{float}
\usepackage[pdftex]{graphicx}
\usepackage{amsmath}
\usepackage{amssymb}

  \usepackage[pdftex,bookmarks,pdfcreator={LaTeX2e, Rainer Kaltenbaek,
	      hyperref}, pdfkeywords={optomechanics, macroscopic, fundamental physics, space, resonator, mechanical, Kaltenbaek, European Space Agency, ESA, quantum mechanics, quantum physics, quantum optics},
	      pdftitle={Testing quantum physics in space using optically trapped nanospheres}, plainpages=false,pdfpagelabels,
	      colorlinks,linkcolor=blue,
	      bookmarksopen,bookmarksopenlevel={0}]{hyperref}

\title{Testing quantum physics in space using optically trapped nanospheres} 

\author{R. Kaltenbaek\supit{a}
\skiplinehalf
\supit{a}Vienna Center for Quantum Science and Technology, Faculty of Physics, University of Vienna, Boltzmanngasse 5, Vienna, Austria
}

\authorinfo{Further author information: E-mail: rainer.kaltenbaek@univie.ac.at, Telephone: 43 1 4277-72534}

\begin{document} 
  \maketitle 

\begin{abstract}
Recent developments in space technology like micro-propulsion systems for drag-free control, thermal shielding, ultra-stable laser sources and stable optical cavities set an ideal platform for quantum optomechanical experiments with optically trapped dielectric spheres. Here, we will provide an overview of the results of recent studies aiming at the realization of the space mission MAQRO to test the foundations of quantum physics in a parameter regime orders of magnitude beyond existing experiments. In particular, we will discuss DECIDE, which is an experiment to prepare and then study a Schr\"o{}dinger-cat-type state, where a dielectric nanosphere of around $100\,$nm radius is prepared in a superposition of being in two clearly distinct positions at the same time. This superposition leads to double-slit-type interference, and the visibility of the interference pattern will be compared to the predictions of quantum theory. This approach allows for testing for possible deviations from quantum theory as our test objects approach macroscopic dimensions. With DECIDE, it will be possible to distinctly test several prominent theoretical models that predict such deviations, for example: the Di\'o{}si-Pensrose model, the continuous-spontaneous-localization model of Ghirardi, Rimini, Weber and Pearle, and the model of K\'a{}rolyh\'a{}zy.  
\end{abstract}



\section{INTRODUCTION}
\label{sec::intro}
The superposition of distinct states is one of the most fundamental concepts of quantum physics. While superposition is well known in the context of (electromagnetic) waves, it is challenging for our understanding in the context of matter waves and, even more so, in the context of superpositions of macroscopically distinct states \cite{Leggett2002a}. That quantum physics allows for an object to simultaneously be in clearly distinct states has been the topic of heated discussions since the early days of quantum physics. Examples are the discussions between Einstein and Bohr \cite{Kumar2009a} regarding the double-slit experiment as well as discussions between Einstein and Schr\"o{}dinger \cite{Prizbram1967a} regarding macroscopic superpositions like Schr\"o{}dinger's cat \cite{Schroedinger1935a}.

By now, the concept of quantum superposition has been confirmed in a series of experiments for increasingly massive objects: for electrons \cite{Davisson1927b,Thomson1927a}, neutrons \cite{Halban1936b}, atoms \cite{Estermann1930a}, and increasingly massive molecules (see, e.g., Refs.~\cite{Arndt1999a,Brezger2002a,Hackermueller2004a,%
Gerlich2007a,Juffmann2012a}). The question remains whether such superpositions can be realized for arbitrarily large and massive objects. Quantum mechanics does not place a principal limitation on macroscopic superpositions. As Schr\"o{}dinger pointed out in his famous gedankenexperiment of Schr\"o{}dinger's cat \cite{Schroedinger1935a}, if one isolates a macroscopic system well enough from its environment, it should, in principle, be possible to realize superpositions of macroscopically clearly distinct states like a cat being dead or alive. Schr\"{o}dinger described the notion of such superpositions as ``burlesk'' \cite{Schroedinger1935a} (``ridiculous'' in the English translation by J. D. Trimmer\cite{Trimmer1980a}). Over the last fifty years, physicists investigated possible modifications of quantum theory that would lead to a rapid ``collapse'' of such macroscopic superpositions. Examples are the model of K\'a{}rolyh\'a{}zy \cite{Karolyhazy1966a,Frenkel1990a}, the models of Di\'o{}si \cite{Diosi1984a,Diosi2007a} and Penrose \cite{Penrose1996a}, the quantum-gravity model of Ellis and co-workers \cite{Ellis1984a} as well as the continuous-spontaneous-localization (CSL) model \cite{Ghirardi1986a,Pearle1976a,Pearle1989a,Gisin1989a,%
Ghirardi1990a,Collett2003a}. Such models are commonly known as macrorealistic models or macrorealistic extensions of quantum theory because they predict a ``realistic'' or ``classical'' behavior of macroscopic objects.

Lab-based matter-wave experiments with increasingly massive objects promise to soon allow for testing certain parameter ranges of the CSL model \cite{Nimmrichter2011b,RomeroIsart2011b,RomeroIsart2011c}. For more ambitious tests of quantum theory over a wider parameter range and/or for decisive tests of the CSL model as well as tests of the models of K\'a{}rolyh\'a{}zy, Di\'o{}si, Penrose or Ellis et al., a laboratory environment is no longer sufficient. In particular, matter-wave interference with increasingly massive objects requires long free-fall times and an exceedingly good micro-gravity environment. For particles with masses on the order of $10^9\,$amu, we expect free-fall times on the order of $100\,$s, which is essentially impossible for ground-based experiments \cite{Kaltenbaek2012b}.

So far, the best option for matter-wave interference with massive objects is Talbot-Lau interferometry. Interferometers of this type have been developed and steadily improved by the Arndt group in Vienna \cite{Arndt1999a,Hackermueller2004a,Juffmann2009a,Gerlich2011a,Juffmann2012a}. In particular, very recently, the Arndt and Hornberger groups presented a new interferometric technique (OTIMA) allowing for matter-wave interferometry with even larger particles \cite{Nimmrichter2011a} than has been possible so far.

While all these experiments were done using ensembles of many particles, recently, it has been proposed to combine the tools of quantum optomechanics and optical trapping in order to cool massive single dielectric particles into the quantum regime \cite{Chang2010a,RomeroIsart2010a,Barker2010a,RomeroIsart2011a}. Building on these proposals, it has now become feasible to perform matter-wave experiments with massive single particles. This approach promises to allow for performing decisive tests of macrorealistic models on Earth \cite{RomeroIsart2011b,RomeroIsart2011c} as well as in space \cite{Kaltenbaek2012b}. The latter is the goal of the potential future space mission MAQRO (macroscopic quantum resonators) and, in particular, of the scientific instrument DECIDE (decoherence in a double-slit experiment) that is to be carried aboard that mission.

\section{MAQRO AND DECIDE}
\label{sec::maqro}
Here, we concentrate on recent results in the development of concepts and technology for MAQRO \cite{Kaltenbaek2012b}. The mission proposal originally comprised two independent experiments: DECIDE and CASE (comparative acceleration sensor). We will concentrate on DECIDE, which implements a double-slit experiment with  dielectric nanospheres with a radius up to $120\,$nm. 

A typical experimental run of DECIDE consists of the following steps:
\begin{enumerate}
\setlength{\itemsep}{0pt}
\setlength{\parskip}{0pt}
\item load a dielectric particle into an optical trap
\item cool the particle's center-of-mass (COM) motion close to the quantum ground state
\item release the particle from the trap (switch off the trapping and cooling fields)
\item let the wavefunction of the particle expand for a time $t_1$
\item switch on a non-linear interaction to prepare the particle in a non-classical Schr\"{o}dinger-cat-like state where the particle is in a superposition of being in two positions ($\pm\Delta x/2$) at the same time.
\item switch off the interaction and let the wavefunction expand for a time $t_2\gg t_1$ sufficient for the two parts of the wavefunction to overlap and interfere
\item switch on the optical field again and measure the position of the particle along the cavity axis via a combination of scattered-light imaging and cavity read out.
\end{enumerate}
The points 2-7 are to be repeated many times in order to acquire enough data points for resolving the interference pattern and determining its visibility. Of course, this requires the dielectric particle to be reusable many times. In other words, the particle should neither get lost during the experiment, nor should it get charged by cosmic or secondary radiation. In a recent ESA study \cite{Kaltenbaek2012a}, we could show that the likelihood for radiation to charge a sub-micron particle is negligible in the case of DECIDE.

In the same study, we investigated in detail the scientific and technical requirements for DECIDE. The scientific requirements are largely determined by the necessary coherence time $t_1+t_2$ and the main decoherence mechanisms limiting the interference visibility according to the predictions of quantum physics:
\begin{itemize}
\setlength{\itemsep}{0pt}
\setlength{\parskip}{0pt}
\item scattering of black-body radiation
\item absorption of black-body radiation
\item emission of black-body radiation
\item scattering of gas molecules
\end{itemize}
The decoherence rates resulting from the first two effects are determined by the environment temperature $T$ while the third effect depends on the internal temperature of the sub-micron sphere, $T_i$. In order for quantum physics to still predict a non-zero interference visibility larger than the visibility allowed for by competing macrorealistic models, one can derive the technical requirements $T \lesssim 16\,$K and $T_i \lesssim 20\,$K.

It should be noted that these requirements have changed since the original proposal\cite{Kaltenbaek2012b} - mainly because the requirements were investigated in more detail in an ESA study concluded in 2012 \cite{Kaltenbaek2012a}. In particular, we found that the expansion time $t_2$ has to be much longer than $t_1$. This has strong consequences for the technical requirements because the various decoherence rates have to be kept even lower than originally expected in order to allow for sufficiently long coherence times $t_1+t_2$ on the order of hundreds of seconds.

\begin{figure}
  \begin{center}
     \begin{tabular}{c}
       \includegraphics[width=0.7\linewidth]{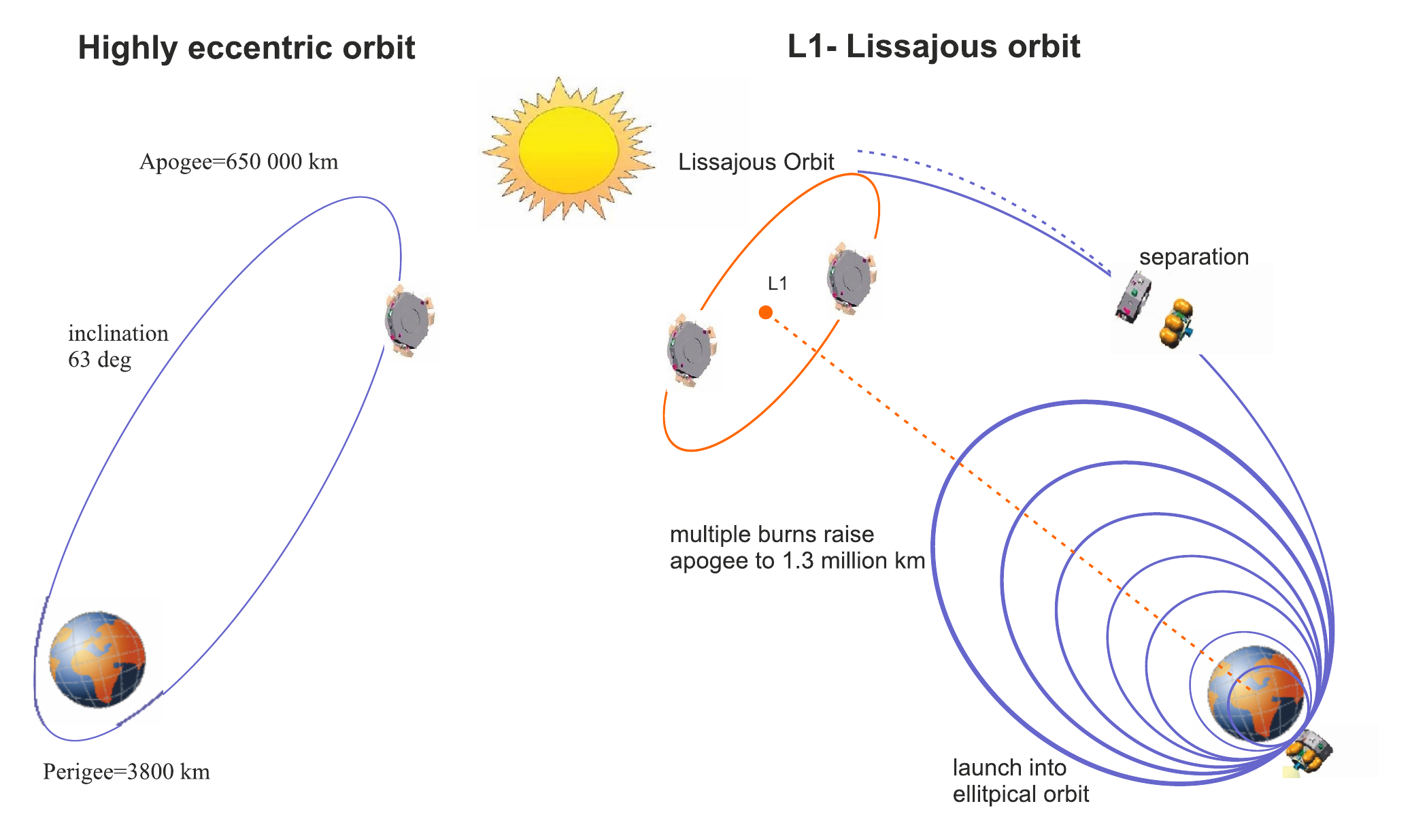}
     \end{tabular}
  \end{center}
  \caption[orbits] 
   { \label{fig::orbits} 
Orbits we considered for MAQRO \cite{Kaltenbaek2012b}. The experiment DECIDE should ideally be performed far from Earth in order to achieve good vacuum levels and low temperatures via passive cooling and direct outgassing into space. In addition, DECIDE requires an excellent microgravity environment. Ideal orbits are, therefore, a Lissajous orbit around L1 or a highly-elliptical orbit, where the experiment could be performed close to the apogee of the orbit. (Graphics by EADS Astrium, based on an image by ESA for the LISA Pathfinder mission)}
   \end{figure} 

Typically, we calculate the influence of decoherence mechanisms in the long-wavelength limit. This limit is applicable for decoherence due to black-body radiation at the low temperatures considered here. It is not applicable for the scattering of (heavy) gas molecules. In order to avoid unnecessary mathematical complications, we set the technical requirement of the gas pressure to $10^{-13}\,$Pa because at such low pressures, the probability for a gas molecule to be scattered off our sub-micron sphere during an experimental run becomes negligible. We expect that a more detailed analysis of decoherence due to gas scattering will result in a more relaxed technical requirement for the gas pressure. However, preliminary estimations of our partners from EADS Astrium show that, for an environment temperature around $16\,$K and direct outgassing into space, gas pressures of $10^{-13}\,$Pa or lower will be achievable using the thermal-shield design of MAQRO. Of course, this requires the spacecraft to be far from Earth's atmosphere, e.g., at the Lagrange point L1 of the Sun-Earth system or at the Apogee of a highly-elliptical orbit (see figure \ref{fig::orbits}).

\section{PROGRESS TOWARDS THE REALIZATION OF DECIDE}
\label{sec::progress}
While the technical requirements for DECIDE are challenging, there has been significant progress towards achieving these requirements and towards increasing the technological readiness level (TRL) of essential technologies that will be required for a future implementation of DECIDE. At the same time, the technologies we are developing for DECIDE, will benefit a wide range of future matter-wave experiments, e.g., for highly sensitive interferometers for Earth observation, gravity-wave detection or for tests of General Relativity.

The main technological challenges for DECIDE have been:
\begin{itemize}
\setlength{\itemsep}{0pt}
\setlength{\parskip}{0pt}
\item the thermal shield (temperatures $\lesssim 16\,$K, pressure $\lesssim 10^{-13}\,$Pa)
\item low-absorption dielectric materials for the optically trapped particle (in order to achieve $T_i \lesssim 20\,$K)
\item free-fall times $t_1$ and $t_2$ and the preparation of a massive dielectric particle in a non-classical Schr\"{o}dinger-cat-like state to realize a double-slit-type matter-wave interference experiment.
\item cavity cooling and feed-back cooling of the center-of-mass motion of a dielectric particle close to the ground state.
\item loading neutral, dielectric sub-micron particles into an optical trap in ultra-high vacuum
\item technical requirements on the micro-propulsion system
\end{itemize}

In the following, we will discuss each of these challenges and describe recent progress aiming at meeting those challenges.

\subsection{The thermal shield}
\begin{figure}
  \begin{center}
     \begin{tabular}{c}
       \includegraphics[width=0.6\linewidth]{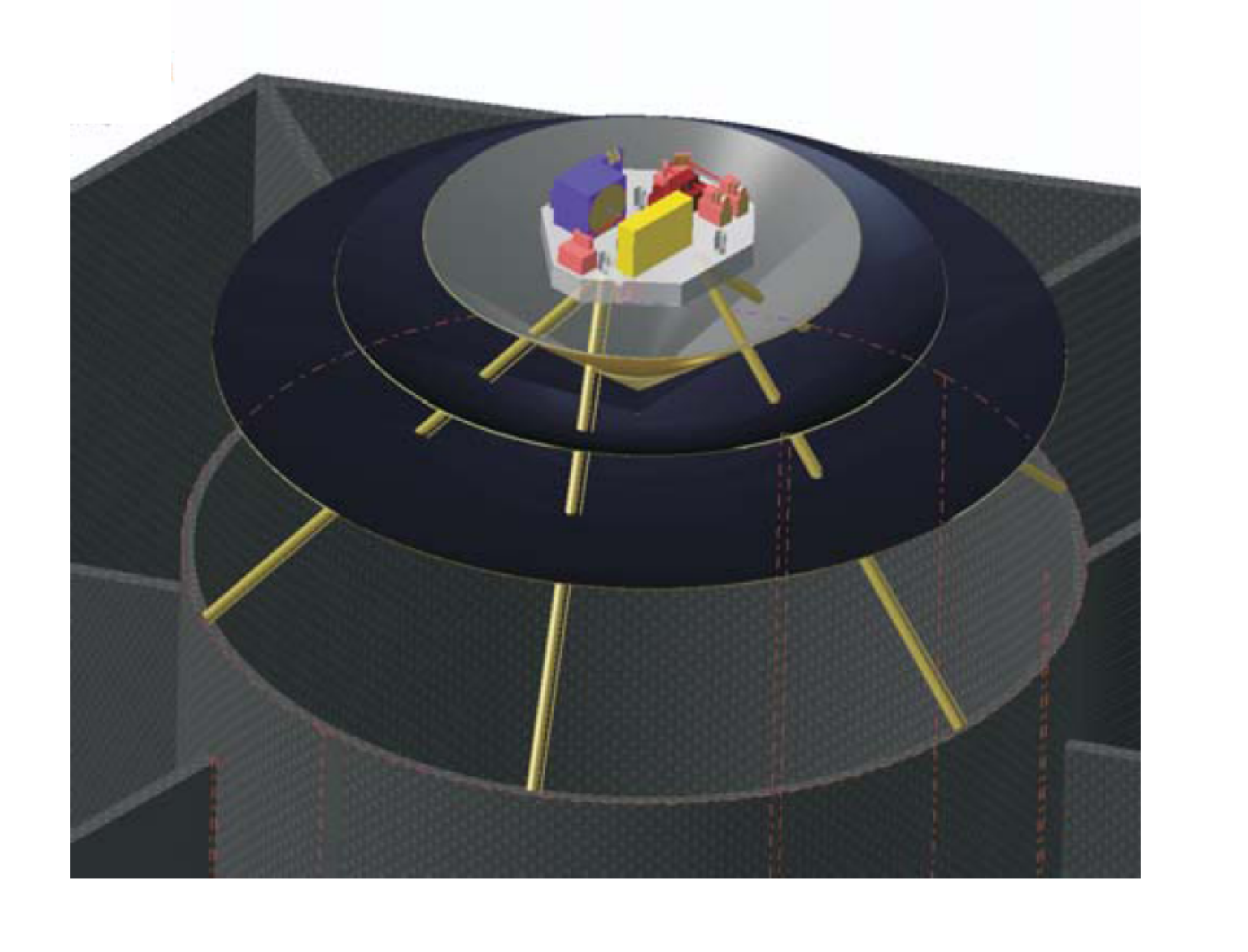}
     \end{tabular}
  \end{center}
  \caption[orbits] 
   { \label{fig::shieldOriginal} 
Design concept for the original design of the thermal shield for MAQRO \cite{Kaltenbaek2012b}. Three shields with increasing opening angles, thermally insulate the optical bench from the hot spacecraft. Thick rods support the structure during launch and can be separated with pyronuts at commission. After commission, the shield assembly is supported by thinner rods to minimize thermal conduction.}
   \end{figure}

Vacuum and cryogenic equipment typically significantly increases the cost of space missions while, at the same time, limiting the mission life time due to limited amounts of cooling agent. Also, active vacuum and cryogenic equipment may be a strong additional source of vibrations. For these reasons, we decided to use an alternative concept for MAQRO based on technological heritage from the GAIA mission and on the mission concept DARWIN. The idea is to place the optical bench for DECIDE outside of the spacecraft and isolate it from the hot spacecraft via a number of thermal shields. Figure \ref{fig::shieldOriginal} shows the original concept for the thermal-shield assembly in the MAQRO mission proposal \cite{Kaltenbaek2012b}.

\begin{figure}
  \begin{center}
     \begin{tabular}{c}
       \includegraphics[width=0.8\linewidth]{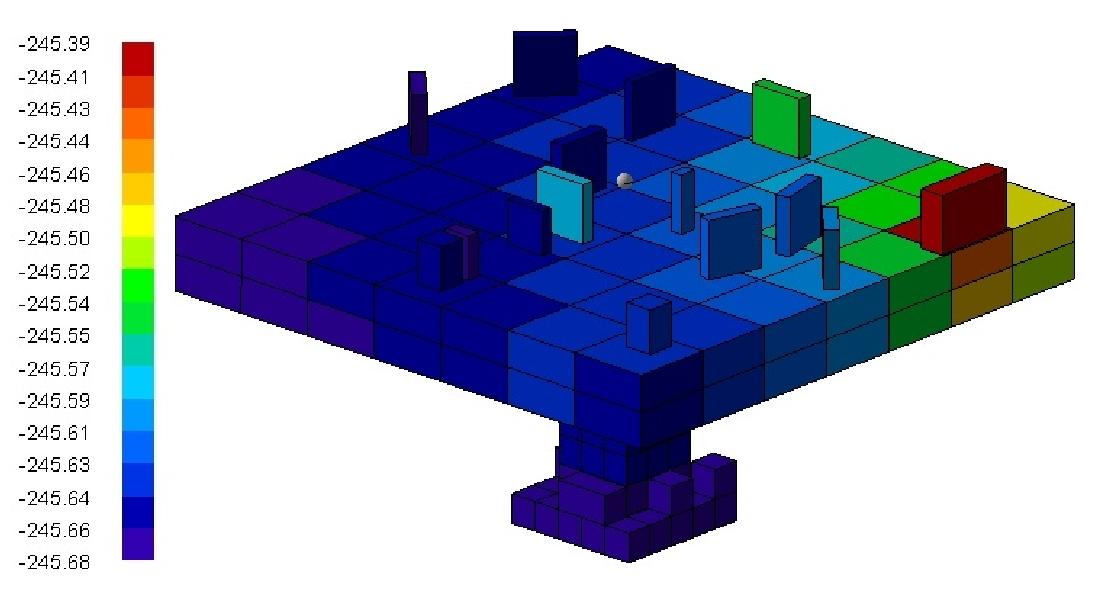}
     \end{tabular}
  \end{center}
  \caption[orbits] 
   { \label{fig::shieldNew} 
Thermal gradients on the optical bench for DECIDE. Picture from the bachelor thesis of F. Hufgard \cite {Hufgard2013a} - results of a detailed thermal analysis performed by F. Hufgard for EADS Astrium in collaboration with the Institute for Space Systems of the University of Stuttgart. The temperatures indicated by the color scale represent the temperatures of various parts of the optical bench and elements on the optical bench in $^\circ$C. A manuscript presenting the results of this thermal study is in preparation. The objects on the bench are mostly mirrors, fiber couplers and a few lenses. The shapes are rough approximations for the thermal analysis. The volume where the dielectric particle will be trapped is simulated by a small, spherical test volume as illustrated in this figure.}
   \end{figure}

Recently, EADS Astrium performed a detailed thermal study of the thermal shield concept \cite{Hufgard2013a}. In this study, various novel developments concerning the optical bench as well as an optimization of the shield parameters were taken into account. In particular, the study investigated how the number of shields, the opening angles and the distances between the shields as well as the introduction of thermally optimized coatings on the shields and on the bench influence the final environment temperature within the region where the dielectric particle is to be trapped for DECIDE. Figure \ref{fig::shieldNew} shows the optimal temperatures achievable on various parts of the optical bench. These temperatures also take into account estimations of the power dissipation of the components on the optical bench. It turned out that the original design of the thermal shield\cite{Kaltenbaek2012b} was close to optimal. In particular, the number of shields (three) has not changed, and the opening angles also are mostly unchanged. Details of the thermal analysis will be published in a manuscript that currently is in preparation.

As can be seen in Figure \ref{fig::shieldNew}, all parts of the optical bench have temperatures $\lesssim 28\,$K. The spherical region where the dielectric particle is trapped has an even lower temperature of $\sim 16.4\,$K, which fulfills the technical requirement for DECIDE. Further improvement of the design like using gold-coated spherical mirrors instead of lenses may further reduce this temperature.
   
\subsection{Free-fall times and preparation of the non-classical state}
\label{subsec::freefall}
The durations of the two free-fall times, $t_1$ and $t_2$, are determined by matter-wave diffraction. If we originally prepare the state of the dielectric particle close to the ground state of the COM motion, the particle's position will be very well defined, and the wavepacket will expand very quickly once the particle is released from the optical trap. For a particle with $10^{-17}\,$kg mass and a angular trapping frequency of $63000\,\mathrm{rad/s}$, the velocity of the wavepacket expansion starting from the ground state is on the order of $1\,\mathrm{\mu m/s}$. For that reason, $t_1$ can be comparatively short (on the order of $1\,$s).

The time $t_2$, on the other hand, is determined by how narrowly we can define the two ``slits'' of our double-slit experiment. This depends on the precise method of how we will prepare the macroscopic superposition. At the moment, we are considering two possibilities:
\begin{enumerate}
\setlength{\itemsep}{0pt}
\setlength{\parskip}{0pt}
\item preparation via post-selection on an $x^2$ position measurement in an auxiliary cavity
\item preparation via local decoherence due to the interaction with a very weak ($10^{-15}\,$W) but narrowly focused laser beam with a wavelength of $\lambda\lesssim 40\,$nm
\end{enumerate}

\subsubsection{Preparation via $x^2$ coupling}
The first method was proposed by O. Romero-Isart and A. Pflanzer et al\cite{RomeroIsart2011b} and investigated in more detail by O. Romero-Isart\cite{RomeroIsart2011c}. The technique is based on using three parallel cavities. After being trapped and cooled in the first cavity, the dielectric particle is released and falls for a time $t_1$. After that time, the particle passes through a second cavity with strong optomechanical coupling. At the same time, a short laser pulse also passes through the cavity. The light reflected from the cavity is used to perform a homodyne measurement in order to read out the position of the passing particle squared ($x^2$). Because the measurement does not reveal the sign of the position, the particle is prepared in a superposition of simultaneously being at positions $\pm x$. The precision of the measurement depends on the coupling strength of the particle to the second cavity. For strong coupling and a large enough value of $x$, the resulting wavepacket is well approximated by two spatially separated narrow Gaussian peaks. The distance of the two peaks is $x$, and the width of the peaks is determined by the coupling strength\cite{RomeroIsart2011b,RomeroIsart2011c}.

After the preparation of this non-classical state, the particle will fall for another time $t_2$ and then enter the third cavity, where the position is measured by a combination of imaging of scattered light and cavity read out.

The advantage of this method is that the two Gaussian peaks can be narrow and well separated. Because the peaks are narrow, diffraction will lead to comparatively quick expansion of the peaks and a correspondingly short time $t_2$ needed for the interference pattern to form (compared to the second method for preparing the non-classical state as discussed below).

One disadvantage is the need for three cavities and the need for the particle to pass from one cavity to the next. In particular, the same particle has to be transported to the first cavity again after each experimental run. Possibilities to overcome this problem are (1) to introduce artificial gravitational gradients on the optical bench, (2) to move the particle via light pressure, and (3) to use one cavity for the preparation and the position measurement and a second, orthogonal  cavity for the preparation of the non-classical state. A second disadvantage is the post selection on the result of the $x^2$ measurement. In principle, the result of the position measurement in the third cavity can be scaled on the measurement result in the second cavity. In this way, it should be possible to use the data gathered in most experimental runs. The scaling may, however, reduce the interference visibility. This will have to be investigated in more detail in the future.

\subsubsection{Preparation via local decoherence}
This is the method we investigated for the original proposal of MAQRO\cite{Kaltenbaek2012b} and for a recently performed ESA study\cite{Kaltenbaek2012a}. The idea is that a narrowly focused, short-wavelength laser beam passes through the volume spanned by the particle wavepacket after the initial expansion time $t_1$. The focused beam is supposed to be smaller than the wavepacket expansion. If the beam is scattered by the particle, the particle will be localized, it the beam misses the particle, the particle will be in a superposition of being anywhere outside the beam path. Because the potentially scattered light is not observed, the particle will then be in a mixture between a well localized state and a Schr\"{o}dinger-cat-like state. After the second time of free fall, $t_2$, the non-classical part of the state will form an interference pattern on top of a broad incoherent background resulting from the localized part of the density matrix.

The advantages of this approach are that (1) the particle can always stay in a single cavity, (2) the method works, in principle, for arbitrarily large particles and arbitrarily large superpositions.

The disadvantages of this approach are that (1) the interference visibility is reduced due to the incoherent background of the localized part of the wavepacket, (2) the two parts of the non-classical state are rather broad and will expand only slowly to overlap and form the interference pattern, (3) the phase between the two interfering parts of the wavepacket is not well defined, which may lead to a reduction of the interference visibility, and (4) the wavelength of the laser beam needed for the preparation of the non-classical state is $\lesssim 40\,$nm, i.e., XUV. In this wavelength regime, the light source and the reflective optics may pose a problem even for the tiny powers (femto-Watt) needed for DECIDE.
   
\subsection{Low-absorption dielectric materials}
The emission of blackbody radiation of a hot particle, is a very strong localization (and, therefore, decoherence) mechanism. Especially for small, sub-micron particles, this has proved to be the dominating decoherence mechanism if the surrounding vacuum is good enough to allow neglecting gas collisions \cite{Kaltenbaek2012a,Kaltenbaek2012b}. We optically trap our particle, and we use optical fields to cool the particle's center-of-mass motion. For that reason, the absorption of the particle's material at the laser wavelength has to be exceedingly low to allow for a low internal particle temperature.

For a laser wavelength of $1064\,$nm, fused silica shows very low absorption down to $0.25\,\mathrm{ppm/cm}$ for Suprasil glass \cite{Hild2006a}. However, in order to achieve the technical requirement of $T_i\lesssim 20\,$K, one would need an absorption of $0.25\,\mathrm{ppb/cm}$\cite{Kaltenbaek2012a} or improved emissivity over the blackbody spectrum. It may be possible to achieve lower absorption coefficient of the dielectric particles by using telecom-wavelength lasers or by using different materials (e.g., pure silicon). Moreover, it will be paramount to take into account the temperature dependence of the optical properties of the particle's material. The requirement on the internal temperature of the particle may also be relaxed by reducing the durations of the free-fall time $t_2$ (see subsection \ref{subsec::freefall}). This is a critical issue that will require further study.

\subsection{Cooling of the COM motion of trapped dielectric particles}
A prerequisite for DECIDE is that the COM motion of the trapped particle is cooled close to the ground state for the motion along the cavity. The COM motion perpendicular to the cavity mode also needs to be cooled in order to prevent coupling between the modes, to limit the read-out noise of the particle position along the cavity, and to prevent loss of the particle out of the cavity mode during the period $t_2$ of free expansion.

In particular, the COM motion along the cavity has to be cooled because, for preparing a double-slit-like state, the initial quantum state has to be spatially coherent over a distance larger than the distance between the two slits. The main reason for cooling the transverse motion of the particle is that the wavefronts of the trapping field  are curved at the trapping position. This leads to an uncertainty in position measurements along the cavity axis. If we want to observe an interference pattern, this uncertainty has to be much smaller than the spacing of the interference fringes.

At the time of submission of the original proposal for MAQRO (2010), neither cavity cooling nor feed-back cooling for sub-micron particles had been observed. Since then, there has been tremendous experimental progress. For example, feed-back cooling of the COM motion of sub-micron particles has been demonstrated by the group of L. Novotny \cite{Gieseler2012a}. Feed-back cooling is especially important for ultra-low pressures where there is not enough gas to cool the motion of the trapped nanosphere against parametric heating, e.g., from laser noise. Feed-back cooling also seems to be essential at modest vacuum levels around $1\,$mbar\cite{Kiesel2013a}. Around that pressure a not-yet-understood heating mechanism results in the loss of optically trapped particles. This can be mitigated by using feed-back cooling\cite{Gieseler2012a}. For lower pressures ($\lesssim 0.1\,$mbar), this heating mechanism seems to be absent, and the particles can be stably trapped without feed-back cooling\cite{Gieseler2013a}. For particles on the micron scale, feed-back cooling was demonstrated by the group of M. Raizen\cite{Li2011a}. Very recently, cavity cooling of sub-micron particles was shown for trapped particles by the Aspelmeyer group\cite{Kiesel2013a,Asenbaum2013a} and for free-flying particles by the Arndt group\cite{Asenbaum2013a}. By combining feed-back and cavity cooling, it seems likely that cooling the COM motion of optically trapped dielectric particles close to the ground state will be achievable in the near future.

\subsection{Loading sub-micron particles into an optical trap}
A central prerequisite for DECIDE is the need for a mechanism to load single, neutral, dielectric particles into an optical trap in high vacuum. First laboratory experiments\cite{Kiesel2013a} currently load the dielectric particles by evaporating a solution containing the particles. Since 2012, we have been working on an experimental study for ESA in order to find more appropriate loading mechanisms for a space environment. This work is performed in collaboration with the Arndt group in Vienna. Preliminary results show that particles of up to $10^8\,$amu can be released. Some of the particles were charged and could be trapped in an ion trap. An alternative approach uses surface-acoustic waves (SAW) induced electronically on a piezoelectric substrate to release particles with $\sim 10^9\,$amu. First results for the SAW approach show that the particles are moving on the substrate but it could not yet be confirmed that they are released. If this can be successfully demonstrated, the SAW technique should, in principle, allow for a quick integration with our existing optical trap, and the SAW devices could be integrated relatively simple with the design of the optical bench of DECIDE.

\subsection{Technical requirements on the micro-propulsion system}
In the recent study we concluded for ESA\cite{Kaltenbaek2012a}, we determined that the requirement on the thruster force noise is $< 1\,\mathrm{\mu N/\sqrt{Hz}}$. This is slightly more relaxed than the requirement for LISA Pathfinder, which is expected to be on the order of $0.1\,\mathrm{\mu N/\sqrt{Hz}}$ \cite{Armano2009}. For a spacecraft with a mass of $700\,$kg, the force-noise requirement for MAQRO corresponds to a requirement on the acceleration noise of $<1.6\times 10^{-9}\mathrm{m/s^2\,Hz^{-1/2}}$.

\section{CONCLUSION AND OUTLOOK}
MAQRO promises to allow expanding the parameter regime over which quantum physics has been tested by several orders of magnitude. At the same time, MAQRO will pave the way for future high-sensitivity matter-wave interferometry in space. Since the original proposal of MAQRO and DECIDE in 2010, there has been significant experimental and theoretical progress towards the realization of MAQRO. In this paper, we have provided an overview over these developments. For many of the technologies and concepts that originally seemed far from the TRL needed for space experiments, there has been significant progress. Based on this rapid development, we are confident regarding the feasibility of realizing DECIDE in the foreseeable future. 

\acknowledgments
We thank G. Hechenblaikner, N. Kiesel, U. Johann, and M. Aspelmeyer for support and stimulating discussions. We thank F. Hufgard for allowing us to use a figure from his bachelor thesis, and we acknowledge financial support from the Austrian Academy of Sciences (APART) and the European Commission (Marie Curie).

\bibliography{/home/rainer/physik/rk.bib}   

\begin{thebibliography}{10}

\bibitem{Leggett2002a}
Leggett, A.~J., ``{T}esting the limits of quantum mechanics: motivation, state
  of play, prospects,'' {\em Journal of Physics: Condensed Matter}~{\bf
  14}(15),  R415 (2002).

\bibitem{Kumar2009a}
Kumar, M.,  [{\em {Q}uantum: {E}instein, {B}ohr and the {G}reat {D}ebate
  {A}bout the {N}ature of {R}eality}{\nolinebreak\hspace{0.1em}]}, Icon Books
  (2000).

\bibitem{Prizbram1967a}
Einstein, A., Schr\"o{}dinger, E., Planck, M., and Lorentz, H.~A., ``{E}instein
  \& {S}chr\"o{}dinger,'' in [{\em {L}etters on wave
  mechanics}{\nolinebreak\hspace{0.1em}]},  Prizbram, K., ed.,  23--42,
  Philosophical Library, New York (1997).

\bibitem{Schroedinger1935a}
Schr{\" o}dinger, E., ``{D}ie gegenw{\" a}rtige {S}ituation in der
  {Q}uantenmechanik,'' {\em {D}ie {N}aturwissenschaften}~{\bf 23}(48),
  807--812 (1935).

\bibitem{Davisson1927b}
Davisson, C. and Germer, L.~H., ``{T}he scattering of electrons by a single
  crystal of nickel,'' {\em Nature}~{\bf 119},  558--560 (1927).

\bibitem{Thomson1927a}
Thomson, G.~P., ``{The Diffraction of Cathode Rays by Thin Films of
  Platinum},'' {\em Nature}~{\bf 120},  802--802 (1927).

\bibitem{Halban1936b}
von Halban, H. and Preiswerk, P., ``{E}xperimental evidence of neutron
  diffraction,'' {\em C.R. Hebd. S\'{e}ances Acad.}~{\bf 203},  73 (1936).

\bibitem{Estermann1930a}
Estermann, I. and Stern, O., ``{B}eugung von {M}olekularstrahlen,'' {\em Z.
  Phys.}~{\bf 61},  95--125 (1930).

\bibitem{Arndt1999a}
Arndt, M., Nairz, O., Voss-Andreae, J., Keller, C., Van~der Zouw, G., and
  Zeilinger, A., ``{W}ave-particle duality of {C}60 molecules,'' {\em
  Nature}~{\bf 401},  680--682 (1999).

\bibitem{Brezger2002a}
Brezger, B., Hackerm{\"u}ller, L., Uttenthaler, S., Petschinka, J., Arndt, M.,
  and Zeilinger, A., ``{Matter-Wave Interferometer for Large Molecules},'' {\em
  Phys. Rev. Lett.}~{\bf 88},  100404 (2002).

\bibitem{Hackermueller2004a}
Hackerm{\"u}ller, L., Hornberger, K., Brezger, B., Zeilinger, A., and Arndt,
  M., ``{D}ecoherence of matter waves by thermal emission of radiatio,'' {\em
  Nature}~{\bf 427},  711--714 (2004).

\bibitem{Gerlich2007a}
Gerlich, S., Hackerm\"uller, L., Hornberger, K., Stibor, A., Ulbricht, H.,
  Gring, M., Goldfarb, F., Savas, T., M{\"u}ri, M., Mayor, M., and Arndt, M.,
  ``{A} {K}apitza-{D}irac-{T}albot-lau interferometer for highly polarizable
  molecules,'' {\em Nature Phys.}~{\bf 3},  711 -- 715 (2007).

\bibitem{Juffmann2012a}
Juffmann, T., Milic, A., M\"{u}llneritsch, M., Asenbaum, P., Tsukernik, A.,
  T\"{u}xen, J., Mayor, M., Cheshnovsky, O., and Arndt, M., ``Real-time
  single-molecule imaging of quantum interference.,'' {\em Nature
  Nanotechnology}~{\bf 7},  297--300 (2012).

\bibitem{Trimmer1980a}
Trimmer, J.~D., ``{T}he {P}resent {S}ituation in {Q}uantum {M}echanics: {A}
  {T}ranslation of {S}chr\"o{}dinger's ``{C}at {P}aradox'','' {\em Proc. Am.
  Phil. Soc}~{\bf 124}(5),  323--338 (1980).

\bibitem{Karolyhazy1966a}
K\'arolyh\'azy, F., ``{G}ravitation and {Q}uantum {M}echanics of {M}acroscopic
  {O}bjects,'' {\em Nuovo Cimento A}~{\bf 52},  390 (1966).

\bibitem{Frenkel1990a}
Frenkel, A., ``{S}pontaneous {L}ocalizations of the {W}ave {F}unction and
  {C}lassical {B}ehavior,'' {\em Found. Phys.}~{\bf 20},  159 (1990).

\bibitem{Diosi1984a}
Di{\'o}si, L., ``Gravitation and quantum-mechanical localization of
  macro-objects,'' {\em Physics Letters A}~{\bf 105}(4--5),  199 -- 202 (1984).

\bibitem{Diosi2007a}
Di{\'o}si, L., ``Notes on certain newton gravity mechanisms of wavefunction
  localization and decoherence,'' {\em Journal of Physics A: Mathematical and
  Theoretical}~{\bf 40}(12),  2989 (2007).

\bibitem{Penrose1996a}
Penrose, R., ``{On Gravity's role in Quantum State Reduction},'' {\em {G}en.
  {R}el. {G}rav.}~{\bf 28},  581--600 (1996).

\bibitem{Ellis1984a}
Ellis, J., Hagelin, J.~S., Nanopoulos, D.~V., and Srednicki, M., ``{Search for
  violations of quantum mechanics},'' {\em Nuclear Physics B}~{\bf 241},
  381--405 (1984).

\bibitem{Ghirardi1986a}
Ghirardi, G.~C., Rimini, A., and Weber, T., ``{U}nified dynamics for
  microscopic and macroscopic systems,'' {\em Phys. Rev.~D}~{\bf 34},  470--491
  (1986).

\bibitem{Pearle1976a}
Pearle, P., ``{R}eduction of the state vector by a nonlinear {S}chr{\"o}dinger
  equation,'' {\em Phys. Rev. D}~{\bf 13},  857--868 (1976).

\bibitem{Pearle1989a}
Pearle, P., ``{C}ombining stochastic dynamical state-vector reduction with
  spontaneous localization,'' {\em Phys. Rev.~A}~{\bf 39},  2277--2289 (1989).

\bibitem{Gisin1989a}
Gisin, N., ``{Stochastic quantum dynamics and relativity},'' {\em Helv. Phys.
  Acta}~{\bf 62},  363--371 (1989).

\bibitem{Ghirardi1990a}
Ghirardi, G.~C., Pearle, P., and Rimini, A., ``Markov processes in hilbert
  space and continuous spontaneous localization of systems of identical
  particles,'' {\em Phys. Rev. A}~{\bf 42},  78--89 (1990).

\bibitem{Collett2003a}
Collett, B. and Pearle, P., ``{Wavefunction Collapse and Random Walk},'' {\em
  Foundations of Physics}~{\bf 33},  1495--1541 (2003).
\newblock 10.1023/A:1026048530567.

\bibitem{Nimmrichter2011b}
Nimmrichter, S., Hornberger, K., Haslinger, P., and Arndt, M., ``Testing
  spontaneous localization theories with matter-wave interferometry,'' {\em
  Phys. Rev. A}~{\bf 83},  043621 (2011).

\bibitem{RomeroIsart2011b}
Romero-Isart, O., Pflanzer, A.~C., Blaser, F., Kaltenbaek, R., Kiesel, N.,
  Aspelmeyer, M., and Cirac, J.~I., ``{Large Quantum Superpositions and
  Interference of Massive Nanometer-Sized Objects},'' {\em Phys. Rev.
  Lett.}~{\bf 107},  020405 (2011).

\bibitem{RomeroIsart2011c}
Romero-Isart, O., ``Quantum superposition of massive objects and collapse
  models,'' {\em Phys. Rev. A}~{\bf 84},  052121 (2011).

\bibitem{Kaltenbaek2012b}
Kaltenbaek, R., Hechenblaikner, G., Kiesel, N., Romero-Isart, O., Schwab,
  K.~C., Johann, U., and Aspelmeyer, M., ``{Macroscopic quantum resonators
  (MAQRO)},'' {\em Experimental Astronomy}~{\bf 34},  123--164 (2012).

\bibitem{Juffmann2009a}
Juffmann, T., Truppe, S., Geyer, P., Mayor, A., Deachapunya, S., Ulbricht, H.,
  and Arndt, M., ``Wave and particle in molecular interference lithography,''
  {\em Phys. Rev. Lett.}~{\bf 103},  263601 (2009).

\bibitem{Gerlich2011a}
Gerlich, S., Eibenberger, S., Tomandl, M., Nimmrichter, S., Hornberger, K.,
  Fagan, P.~J., T\"{u}xen, J., Mayor, M., and Arndt, M., ``{Quantum
  interference of large organic molecules.},'' {\em Nature Communications}~{\bf
  2},  263 (2011).

\bibitem{Nimmrichter2011a}
Nimmrichter, S., Haslinger, P., Hornberger, K., and Arndt, M., ``Concept of an
  ionizing time-domain matter-wave interferometer,'' {\em New Journal of
  Physics}~{\bf 13}(7),  075002 (2011).

\bibitem{Chang2010a}
Chang, D.~E., Regal, C.~A., Papp, S.~B., Wilson, D.~J., Ye, J., Painter, O.,
  Kimble, H.~J., and Zoller, P., ``Cavity opto-mechanics using an optically
  levitated nanosphere.,'' {\em Proceedings of the National Academy of Sciences
  of the United States of America}~{\bf 107}(3),  1005--1010 (2010).

\bibitem{RomeroIsart2010a}
Romero-Isart, O., Juan, M.~L., Quidant, R., and Cirac, J.~I., ``{T}oward
  quantum superposition of living organisms,'' {\em New J. Phys.}~{\bf 12},
  033015 (2010).

\bibitem{Barker2010a}
Barker, P.~F. and Shneider, M.~N., ``Cavity cooling of an optically trapped
  nanoparticle,'' {\em Phys. Rev. A}~{\bf 81},  023826 (2010).

\bibitem{RomeroIsart2011a}
Romero-Isart, O., Pflanzer, A.~C., Juan, M.~L., Quidant, R., Kiesel, N.,
  Aspelmeyer, M., and Cirac, J.~I., ``Optically levitating dielectrics in the
  quantum regime: Theory and protocols,'' {\em Phys. Rev. A}~{\bf 83},  013803
  (2011).

\bibitem{Kaltenbaek2012a}
Kaltenbaek, R., Hechenblaikner, G., et~al., ``Macroscopic quantum experiments
  in space using massive mechanical resonators,'' tech. rep., Study conducted
  under contract with the European Space Agency, Po P5401000400 (2011--2012).

\bibitem{Hufgard2013a}
Hufgard, F., ``{Thermal Shield Design for Mission ``Macroscopic Quantum
  Oscillators in Space'' },'' (2013).

\bibitem{Hild2006a}
Hild, S., L\"{u}ck, H., Winkler, W., Strain, K., Grote, H., Smith, J., Malec,
  M., Hewitson, M., Willke, B., Hough, J., and Danzmann, K., ``Measurement of a
  low-absorption sample of oh-reduced fused silica,'' {\em Appl. Opt.}~{\bf
  45},  7269--7272 (2006).

\bibitem{Gieseler2012a}
Gieseler, J., Deutsch, B., Quidant, R., and Novotny, L., ``{Subkelvin
  Parametric Feedback Cooling of a Laser-Trapped Nanoparticle},'' {\em Phys.
  Rev. Lett.}~{\bf 109},  103603 (2012).

\bibitem{Kiesel2013a}
Kiesel, N., Blaser, F., Delic, U., Grass, D., Kaltenbaek, R., and Aspelmeyer,
  M., ``{Cavity cooling of an optically levitated nanoparticle},''  14 (2013).
\newblock arXiv:1304.6679.

\bibitem{Gieseler2013a}
Gieseler, J., Novotny, L., and Quidant, R., ``{Thermal nonlinearities in a
  nanomechanical oscillator},''  12 (2013).
\newblock arXiv:1307.4684.

\bibitem{Li2011a}
Li, T., Kheifets, S., and Raizen, M.~G., ``{Millikelvin cooling of an optically
  trapped microsphere in vacuum},'' {\em Nature Physics}~{\bf 7},  527--530
  (2011).

\bibitem{Asenbaum2013a}
Asenbaum, P., Kuhn, S., Nimmrichter, S., Sezer, U., and Arndt, M., ``{Cavity
  cooling of free silicon nanoparticles in high-vacuum},'' (2013).
\newblock arXiv:1306.4617.

\bibitem{Armano2009}
M.~Armano, e.~a., ``{LISA Pathfinder: The experiment and the route to LISA},''
  {\em Class. Quantum Grav.}~{\bf 26}(9),  094001 (2009).

\end{thebibliography}
\bibliographystyle{spiebib}   

\end{document}